\pdfoutput=1
\documentclass[ ]{svjour3}

\usepackage{graphicx,color}

\begin{document}

\title{Electronic and magnetic nano phase separation in cobaltates La$_{2-x}$Sr$_{x}$CoO$_4$}

\author{Z. W. Li, Y. Drees, A. Ricci, D. Lamago, A. Piovano, M. Rotter, W. Schmidt, O. Sobolev, U. R\"{u}tt, O. Gutowski, M. Sprung, J. P. Castellan, L. H. Tjeng \and A.~C.~Komarek*}

\authorrunning{Z. W. Li, ... and A.~C.~Komarek*} 

\institute{Alexander Christoph Komarek*, Zhiwei Li, Yvo Drees, Martin Rotter
and Liu Hao Tjeng \at
              Max-Planck-Institute for Chemical Physics of Solids, N\"{o}thnitzer Str. 40, D-01187 Dresden, Germany \\
              \email{Alexander.Komarek@cpfs.mpg.de}           
\and
Alessandro Ricci, Uta R\"{u}tt, Olof Gutowski and Michael Sprung
\at
              Deutsches Elektronen-Synchrotron DESY, Notkestr. 85, 22603 Hamburg, Germany
\and
Daniel Lamago and John-Paul Castellan \at
              Laboratoire L\'{e}on Brillouin, CEA/CNRS,F-91191 Gif-sur Yvette Cedex, UMR12 CEA-CNRS, B\^{a}t 563 CEA Saclay,
              France\\
              Institute of Solid State Physics, Karlsruhe Institute of Technology, D-76021 Karlsruhe,
              Germany\\
\and
Andrea Piovano \at
              Institut Laue-Langevin (ILL), 6 Rue Jules Horowitz, F-38043 Grenoble, France
\and
Wolfgang Schmidt     \at
              J\"{u}lich Centre for Neutron Science JCNS, Forschungszentrum J\"{u}lich GmbH, Outstation at ILL, CS 20156, 71 avenue de Martyrs, 38042 Grenoble, France
\and
Oleg Sobolev \at
              Forschungsneutronenquelle Heinz Maier-Leibnitz (FRM-II), TU M\"{u}nchen, Lichtenbergstr. 1, D-85747 Garching,
              Germany\\
              Georg-August-Universität G\"{o}ttingen, Institut für Physikalische Chemie, Tammannstrasse 6, D-37077 G\"{o}ttingen,
              Germany
}
\date{Received: date / Accepted: date}

\maketitle

\begin{abstract}
The single-layer perovskite cobaltates have attracted
enormous attention due to the recent observation of hour-glass shaped
magnetic excitation spectra which resemble the ones of the famous high-temperature superconducting
cuprates. Here, we present an overview of our most recent studies of the spin and charge
correlations in floating-zone grown cobaltate single crystals.
We find that frustration and a novel kind of electronic and magnetic nano phase separation are
intimately connected to the appearance of the hour-glass shaped spin excitation spectra. We also point out the difference between nano phase separation and conventional phase separation.
\end{abstract}

\section{Introduction}
\label{intro}
The hour-glass magnetic excitation spectrum has
fascinated physicists over years. This spectrum has been observed in
high-temperature superconducting cuprates and it is widely believed
that fluctuating charge stripes are involved in the physics of these
intriguing materials \cite{tranquada,vojta}. Besides charge stripes
also Fermi surface effects have been proposed for the emergence of
these hour-glass spectra in cuprates\cite{eremin}. Recently,
Boothroyd \emph{et al.} observed an hour-glass magnetic spectrum in
the copper-free Co oxide material La$_{2-x}$Sr$_{x}$CoO$_4$ implying
that Fermi surface effects are not needed for the emergence of
hour-glass spectra since this material is insulating
\cite{boothroyd}. The undoped parent material La$_{2}$CoO$_4$ is an
antiferromagnetic insulator with a N\'{e}el-temperature of roughly
275~K \cite{babkevich}. In contrast to the high-temperature
superconducting cuprates La$_{2-x}$Sr$_{x}$CuO$_4$, the cobaltates
remain insulating up to high hole-doping levels of $x$~$\sim$~1
\cite{moritomo}. Also the antiferromagnetic correlations with
La$_{2}$CoO$_4$-like character which are centered at the planar
antiferromagnetic wavevector persist to very high hole-doping ranges
\cite{dreesA}, higher than in the cuprates where incommensurate
magnetic satellites have been observed already above $\sim$2\%
Sr-doping \cite{matsuda}. A clearly incommensurate antiferromagnetic
regime with well-separated incommensurate antiferromagnetic peaks
can be observed only above  $\sim$33\% of hole-doping in the
cobaltates \cite{dreesA}. At half-doping a robust checkerboard
charge order (CBCO) has been reported in these cobaltates
(La$_{1.5}$Sr$_{0.5}$CoO$_4$) with a charge ordering temperature of
the order of $\sim$800~K \cite{zaliznyak,helme}. It was shown that
the Co$^{2+}$ and Co$^{3+}$ ions in La$_{1.5}$Sr$_{0.5}$CoO$_4$ are
in the high spin and in the nonmagnetic low-spin state respectively
\cite{helme,chang}.
\begin{figure}[!b]
  \includegraphics[width=1\textwidth]{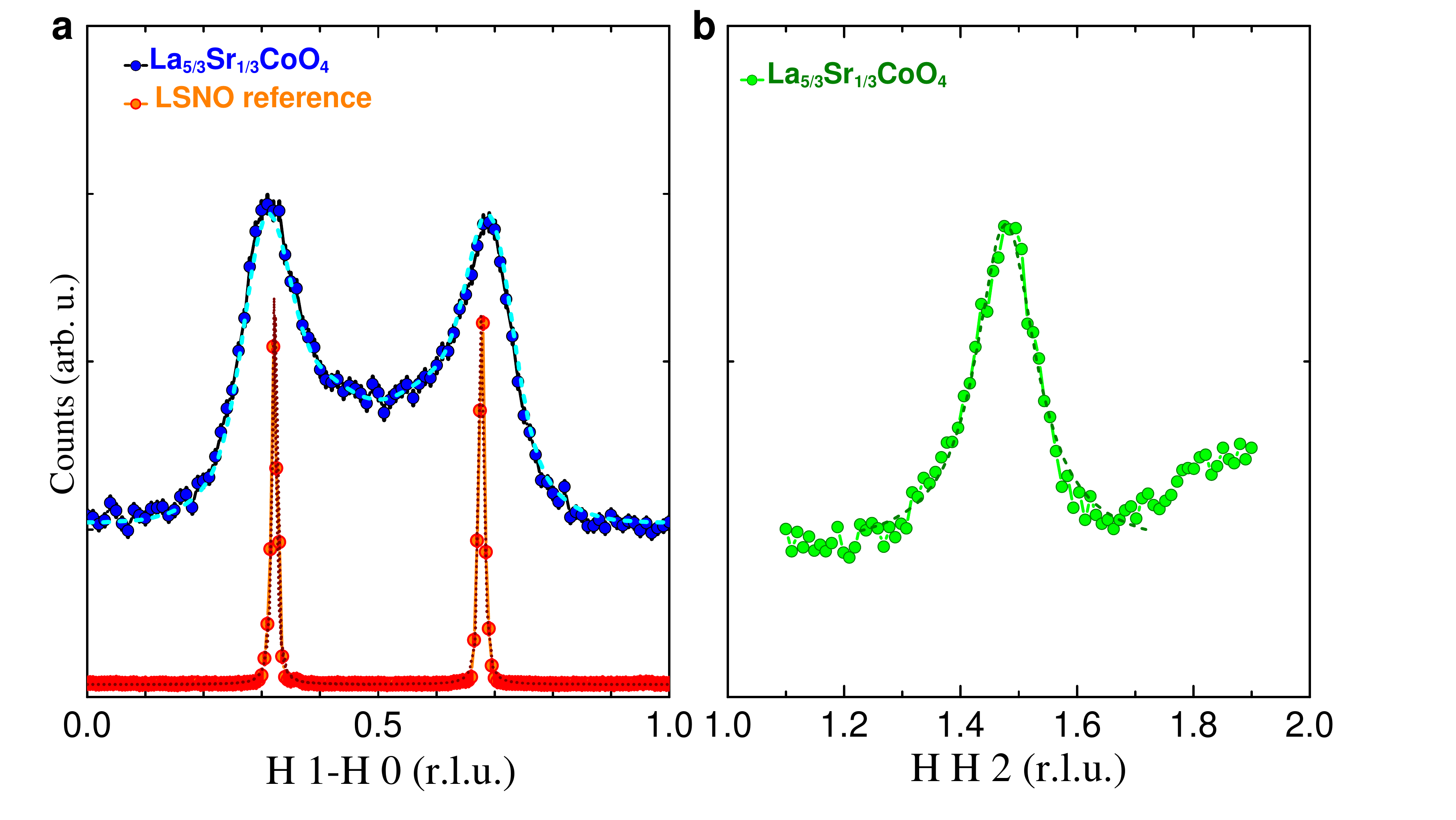}
\caption{\textbf{a} Magnetic neutron scattering intensities at low
temperatures for one-third hole-doped cobaltate and nickelate
respectively \cite{dreesB}. \textbf{b} The charge correlations
observed at 285~K in La$_{5/3}$Sr$_{1/3}$CoO$_4$ \cite{dreesB}.}
\label{fig1}
\end{figure}

\section{Results and Discussion}

At one-third doping, i.e. in La$_{5/3}$Sr$_{1/3}$CoO$_4$, broad
incommensurate magnetic peaks have been observed around
third-integer positions in reciprocal space. Also in the
isostructural nickelates La$_{2-x}$Sr$_{x}$NiO$_4$ \cite{tranquadaB}
magnetic peaks can be observed at about the same positions in
reciprocal space around 1/3-hole-doping, see Fig.~\ref{fig1}~(a).
These nickelates are known to exhibit a robust diagonal charge
stripe phase \cite{tranquadaB}. Hence, it is natural to expect the
existence of charge stripe ordered phases also in
La$_{5/3}$Sr$_{1/3}$CoO$_4$, with the note that the correlation
length of these diagonal charge stripes was thought to be very low
in the cobaltate corresponding to the broadness of the magnetic
\begin{figure}[!b]
  \includegraphics[width=1\textwidth]{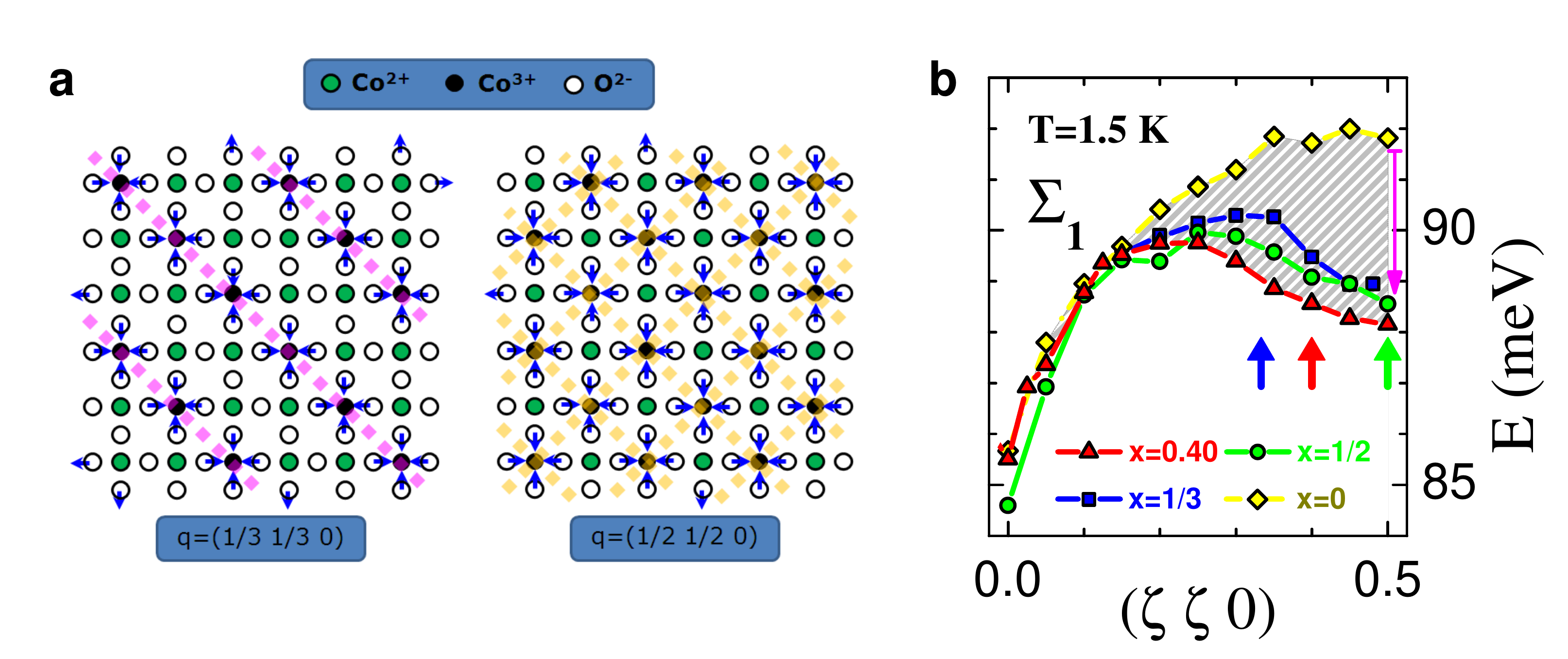}
\caption{\textbf{a} Polarization patterns for two high-frequency
$\Sigma_1$ Co-O bond-stretching phonon dispersions with propagation
vectors of (1/3~1/3~0) and (1/2~1/2~0) respectively are shown. As
can be seen, one would expect a coupling of a phonon mode to charge
correlations that have the same propagation vector as the phonon
mode. \textbf{b} The Co-O bond-stretching phonon dispersions of
La$_{2-x}$Sr$_{x}$CoO$_4$ with $x$~=~0 (yellow), 1/3 (blue), 0.4
(red) and 1/2 (green) \cite{dreesA}. For a better comparison, the
dispersion of  La$_{2}$CoO$_4$ has been shifted to higher energies.
The blue/red/green arrow marks the position of the propagation
vector of a 1/3-doped and of a 40\% hole-doped charge stripe phase
and of a CBCO phase. As can bee seen,  La$_{5/3}$Sr$_{1/3}$CoO$_4$
exhibits absolutely no anomalies in the phonon dispersion at
propagation vectors corresponding to a 1/3-hole-doped charge stripe
phase (blue arrow). Instead, a phonon softening occurs at the zone
boundary which is indicative for CBCO. Also
La$_{1.6}$Sr$_{0.4}$CoO$_4$ exhibits a similar phonon softening
(magenta arrow) like the half-doped CBCO ordered cobaltate
La$_{1.5}$Sr$_{0.5}$CoO$_4$. }
\label{fig2}       
\end{figure}
peaks. However, using 100~keV hard X-rays at the synchrotron we were
able to show that in La$_{5/3}$Sr$_{1/3}$CoO$_4$ the charge
correlations essentially occur with CBCO character only, see
Fig.~\ref{fig1}~(b) \cite{dreesB}. Also in La$_{1.6}$Sr$_{0.4}$CoO$_4$ the charge
ordering peaks were not observed at the incommensurate positions
that correspond to the magnetic peak positions, but, instead at
half-integer peak positions typical for CBCO \cite{dreesA,guo}. The CBCO
character of the charge correlations in La$_{5/3}$Sr$_{1/3}$CoO$_4$
and La$_{1.6}$Sr$_{0.4}$CoO$_4$ is also reflected in the topmost
$\Sigma_1$ Co-O bond-stretching phonon dispersions that resemble the
one in the half-doped CBCO ordered cobaltate
La$_{1.5}$Sr$_{0.5}$CoO$_4$ where an anomalous phonon softening can
be observed at half-integer propagation vectors, see magenta arrow
in Fig.~\ref{fig2}. This is unlike to the behaviour in the undoped
parent material La$_{2}$CoO$_4$ and also unlike to the behaviour in
charge stripe ordered nickelates \cite{dreesB}.
\par Interestingly, nickelates around half-doping also exhibit CBCO at high temperatures ($\sim$480 K) and become charge stripe ordered on cooling to low temperatures ($\sim$180 K) \cite{kajimoto,ishizaka}.
This re-arrangement of charges into stripes at low temperatures seems to be hampered in the cobaltates which also have a much more robust CBCO with distincly higher CBCO ordering temperatures than in the nickelates.
\begin{figure}[!b]
  \includegraphics[width=0.67\textwidth]{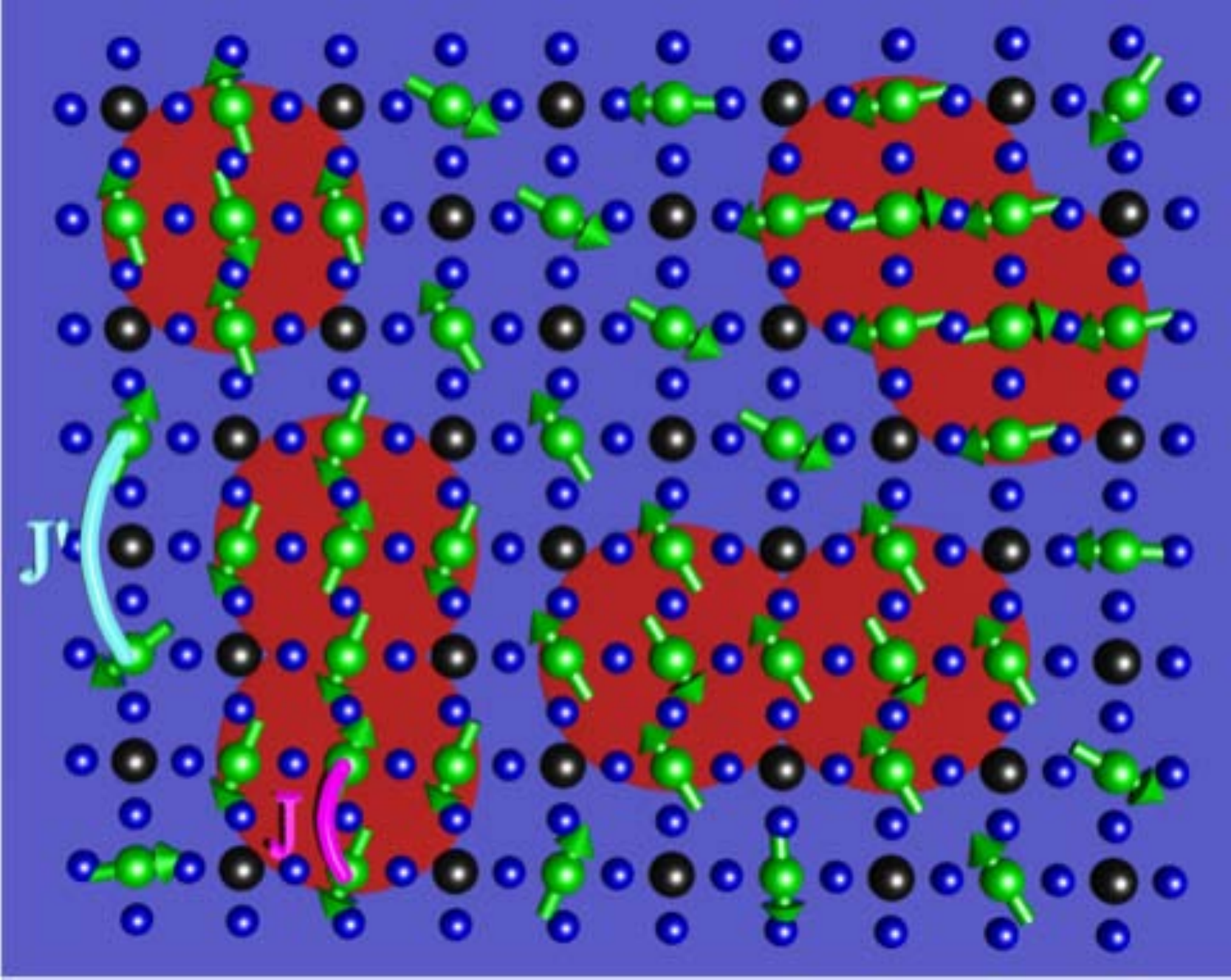}
\caption{The Co$^{2+}$ ions (green speheres with arrows symbolizing
the spin) and the nonmagnetic Co$^{3+}$ ions (black spheres) within
the $ab$-plane are schematically drawn for our nano phase separation
scenario. By doping of electrons at Co$^{3+}$-sites in a
checkerboard charge ordered matrix (blue) undoped islands (red) will
be created. Within the undoped islands the nearest-neighbour
exchange interactions $J$ are quite large. However, in the
surrounding CBCO regions only weak exchange iteractions $J'$ across
the holes exist \cite{guo}.}
\label{fig3}
\end{figure}
An apparent difference between cobaltates and nickelates is the higher insulating properties of cobaltates according to the spin-blockade mechanism \cite{chang}
as well as the stability of both Co$^{2+}$ and Co$^{3+}$ oxidation states (charges are localized at Co-sites and not at the oxygen ions) together with a very large difference in their ionic sizes (the ratio of the Co$^{2+}$ and Co$^{3+}$  ionic radii amounts to $\sim$1.367 \cite{shannon}). All these effects together hamper the hopping of electrons from Co$^{2+}$ to Co$^{3+}$ sites and, thus, probably the re-arrangement of charges into stripes at low temperatures.
\par The absence of charge stripes and Fermi surfaces in La$_{5/3}$Sr$_{1/3}$CoO$_4$ and La$_{1.6}$Sr$_{0.4}$CoO$_4$ forced us to find an alternative mechanism for the emergence of the hour-glass spectrum \cite{dreesA,dreesB}.
We proposed that frustration effects in disordered CBCO phases produce the incommensurabilities in La$_{2-x}$Sr$_{x}$CoO$_4$ \cite{dreesA}.
Even more intriguing is the microscopic origin behind this mechanism. For cobaltates below half-doping additional electrons have to be introduced into the Co-oxygen planes, i.e. Co$^{2+}$ ions have to be inserted at nonmagnetic Co$^{3+}$ sites. This electron doping of CBCO phases will create undoped La$_{2}$CoO$_4$-like islands on the nanometer scale. Within these undoped islands large nearest-neighbour exchange interactions $J$ occur, whereas the surrounding regions with CBCO character only contain much weaker exchange interactions $J'$ accross the holes, see Fig.~\ref{fig3}. This gives rise to a new situation - whereas nanometer-sized undoped islands can be excited to high energies the surrounding CBCO ordered regions are not able to follow these high-frequency oscillations.
\begin{figure}[!b]
  \includegraphics[width=1\textwidth]{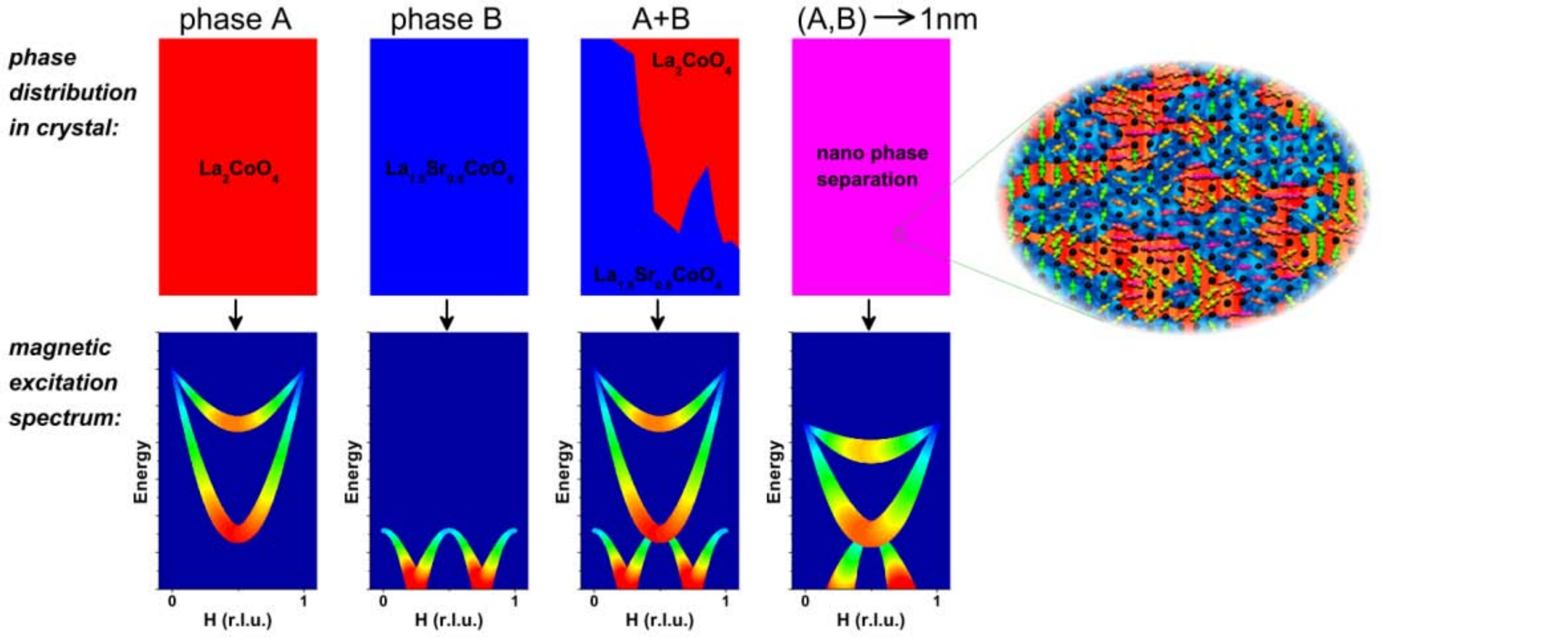}
\caption{The first two columns schematically show the magnetic
excitation spectra of the undoped parent material La$_{2}$CoO$_4$
(phase A) and of the checkerboard charge ordered half-doped material
La$_{1.5}$Sr$_{0.5}$CoO$_4$ (phase B). The reported \cite{helme}
extremely weak optical mode in the spectrum of
La$_{1.5}$Sr$_{0.5}$CoO$_4$ has been omitted since it could not be
detected with a triple axis-spectrometer in Ref.~\cite{dreesA} in
the accessible Q-space and can be neglected for these reasons). The
third column schematically shows the expected observations for
conventional phase separation with any domain sizes down to the
micro-meter scale very roughly. Finally, the last column
schematically shows the situation for nano phase separation. When
domain sizes shrink to the nano scale the magnetic spectra are no
longer a trivial superposition of the spectra of phases A and B.
Instead, a novel spectrum with the characteristics of an hour-glass
spectrum appears. Thus, nano phase separation is distinct from
conventional phase separation and might induce new physical
properties.}
\label{fig4}
\end{figure}
Hence, besides electronic (i.e. charge) also magnetic nano phase separation occurs in these cobaltates.
The high energy excitations within the hour-glass shaped magnetic spectra appear to be basically hosted in nano phase separated undoped islands whereas the remaining hole-rich regions can be excited only at lower energies such that the so-called hour-glass spectrum is formed from the coupling of these two very different type of excitations.
\par The magnetic part of our nano phase separation scenario was corroborated by our study of the hole-doping dependence and the temperature dependence of the hour-glass spectra \cite{dreesB}. The charge part of our nano phase separation scenario was also verified by microdiffraction measurements on La$_{5/3}$Sr$_{1/3}$CoO$_4$ at the synchrotron,
where we scanned a CBCO peak reflection intensity with a
micro-focused beam across the surface. The domain size distribution
exhibits islands with larger and islands with smaller domain sizes
(correlation lengths). Using the statistical method developed in
Ref.~\cite{fratini} we analyzed the probability density function
(PDF) and observed an almost perfect power-law behaviour that
revflects a scale-invariant (fractal-like) domain size distribution
which nicely confirms our nano phase separation scenario in
cobaltates \cite{dreesB}.
Using these new tools it has been possible to discover a common
(structural) nanoscale phase separation scenario characterized by
the coexistence of competing granular networks of different striped
orders
\cite{fratini,pocciaA,pocciaB,pocciaC,pocciaD,ricciA,ricciB,innocenti,ricciC,campiD,ricciE,ricciF,pocciaE,pocciaF,ricciG,ricciH}.
However, so far a rigorous study of the dynamical interplay among
these multiple orders is still missing but a perspective might come
from the use of the coherent x-ray beams available at the new
generation synchrotron facilities and free electron lasers
\cite{ricciI}.
\par Here, we would like to emphasize that nano phase separation appears to be different from conventional phase separation as is schematically shown in Fig.~\ref{fig4}.
Whereas conventional phase separation will basically lead to a trivial superposition of magnetic excitation spectra of the contributing phases,
electronic and magnetic nano phase separation is able to create new physical properties. As is shown schematically in Fig.~\ref{fig4} the energy scale and width of the magnetic excitations are altered and even the entire spectrum looks different from a simple superposition of the spectra of the contributing phases since the outwards-dispersing branches are suppressed in the low energy part of the spectrum.
Whereas, the bulk properties of a material are usually governed by the volume properties of the contributing phases if conventional phase separation occurs,
it is this volume of the contributing phases which has almost vanished as soon as nano phase separation occurs. This is, then, responsible for new physical properties as is shown for the example of the hour-glass shaped magnetic excitation spectra in cobaltates.

\begin{acknowledgements}
A.~C.~K. thanks J. Zaanen and G. Seibold for valuable discussions.
\end{acknowledgements}

\end{document}